\documentclass[manuscript]{aastex}

\shorttitle{Pluto and Triton}
\shortauthors{Tegler et al.}

\begin{document}

%Version March 16, 2012

\title{Ice Mineralogy Across and Into the Surfaces of Pluto, Triton, and Eris}


\author{S. C. Tegler\altaffilmark{1} }
\affil{Department of Physics and Astronomy, Northern Arizona University,
    Flagstaff, AZ 86011}
\email{Stephen.Tegler@nau.edu}

\author{W. M. Grundy\altaffilmark{2}}
\affil{Lowell Observatory,  Flagstaff, AZ  86001 }
\email{W.Grundy@lowell.edu}

\author{C. B. Olkin\altaffilmark{2}}
\affil{Southwest Research Institute,  Boulder, CO  80302}
\email{colkin@boulder.swri.edu}

\author{L. A. Young\altaffilmark{2}}
\affil{Southwest Research Institute,  Boulder, CO 80302}
\email{layoung@boulder.swri.edu}

\author{W. Romanishin\altaffilmark{1}}
\affil{Department of Physics and Astronomy, University of Oklahoma, Norman, OK  73019}
\email{wjr@nhn.ou.edu}

\author{D.M. Cornelison\altaffilmark{}}
\affil{Department of Physics, Astronomy, and Materials Science, Missouri State University, Springfield, MO 65897}
\email{DavidCornelison@MissouriState.edu}

\author{R. Khodadadkouchaki\altaffilmark{}}
\affil{Department of Physics, California State University, Northridge, CA 91330}


\altaffiltext{1}{Visiting Astronomer, MMT Observatory. Observations reported here were obtained
at the MMT Observatory, a joint facility of the University of Arizona and the Smithsonian Institution }
\altaffiltext{2}{Visiting Astronomer, NASA IRTF}
%\altaffiltext{3}{Current address: Department of Physics, California State University, Northridge, CA 91330, USA}
%\altaffiltext{3}{Present Address: Department of Physics, Astronomy, and Materials Science, Missouri State University, Springfield, MO, 65804 }

\begin{abstract}
We present three near-infrared spectra of Pluto taken with the IRTF and SpeX, an optical spectrum of Triton taken with the MMT and the Red Channel Spectrograph, and previously published spectra of Pluto,  Triton, and Eris. We combine these observations with a two-phase Hapke model, and gain insight into the ice mineralogy on Pluto, Triton, and Eris.  Specifically, we measure the methane-nitrogen mixing ratio across and into the surfaces of these icy dwarf planets.
In addition, we present  a laboratory experiment that demonstrates it is essential to model methane bands in spectra of icy dwarf planets with two methane phases $-$ one highly-diluted by nitrogen and the other rich in methane. 

For Pluto, we find bulk, hemisphere-averaged, methane abundances of 9.1 $\pm$ 0.5\%, 7.1 $\pm$ 0.4\%, and 8.2 $\pm$ 0.3\% for sub-Earth longitudes of 10$^{\circ}$, 125$^{\circ}$, and 257$^{\circ}$. Application of the Wilcoxon rank sum test to our measurements finds these small differences are statistically significant. For Triton, we find bulk, hemisphere-averaged, methane abundances of 5.0 $\pm$ 0.1\% and 5.3 $\pm$ 0.4\% for sub-Earth longitudes of 138$^{\circ}$ and 314$^{\circ}$. Application of the Wilcoxon rank sum test to  our measurements finds the differences are not statistically significant. For Eris, we find  a bulk, hemisphere-averaged, methane abundance of 10 $\pm$ 2\%.

Pluto, Triton, and Eris do not exhibit a trend in methane-nitrogen mixing ratio with depth into their surfaces over the few cm range probed by these observations. This result is contrary to the expectation \citep{gru00} that since visible light
penetrates deeper into a nitrogen-rich surface than the depths from which thermal emission emerges, net radiative heating at depth would drive preferential sublimation of nitrogen leading to an increase in the methane abundance with depth.
\end{abstract}


\keywords{methods: laboratory --- methods: observational --- planets and satellites: surfaces --- techniques: spectroscopic}



\section{Introduction}
By comparing spectra of icy dwarf planets taken with telescopes and spectra of ice samples taken in the laboratory, it is possible to gain considerable insight into the ice mineralogy of these objects.  Such insight helps identify important formation and evolution mechanisms at work in the outer Solar System. It also provides context for the New Horizons spacecraft flyby of Pluto in 2015.

This paper follows on earlier work  \citep{teg10} in which we compared optical spectra of Pluto and Eris with laboratory spectra of methane and nitrogen ice samples in order to derive methane and nitrogen abundances for these objects. We found bulk, hemisphere-averaged, methane and nitrogen mixing ratios of $\sim3\%$ and $\sim97\%$ for Pluto and $\sim10\%$ and $\sim90\%$ for Eris. 

Those Eris measurements challenged previous work including our own that had concluded nitrogen was a
minority species rather than a majority species \citep{bro05,dum07,abe09,mer09}. That conclusion rested on two observations. First, the nitrogen 2.15 $\mu$m band has yet to be detected in spectra of Eris, but is seen in spectra of Pluto. Second, methane ice bands in spectra of Eris show smaller blue shifts away from the position of pure-methane ice absorption than  bands in spectra of Pluto.  Laboratory experiments show that methane ice bands shift to shorter wavelengths when methane is highly-diluted with nitrogen \citep{qui97}, and the larger the nitrogen content the larger the shift \citep{bru08}. 

We arrived at the result in that paper by taking the nitrogen-methane binary phase diagram generated from x-ray diffraction studies into account \citep{py83}. The binary phase diagram indicates that at temperatures relevant to the surfaces of icy dwarf planets, only small amounts of methane are soluble in nitrogen ice, and only small amounts of nitrogen are soluble in methane ice. For bulk compositions between the solubility limits (which are strong functions of temperature), thermodynamic equilibrium dictates that two phases must coexist: methane ice saturated with nitrogen and nitrogen ice saturated with methane. For a given bulk composition and temperature, the phase diagram shows how much of each phase must be present.  \cite{ls85} were the first to suggest  a physical state for volatiles on icy dwarf planets in agreement with the above discussion. Specifically, they suggested small amounts of methane were dissolved in nitrogen ice on the surface of Triton.  By including two phases in our model of Eris' methane ice bands, we arrived at our result for the methane-nitrogen mixing ratio.   

Our work on Eris departed from previous work in another way. We found no evidence for a trend in the methane-nitrogen mixing ratio with depth into the surface of Eris whereas previous work found evidence for a trend.  Previous work used the observed blue shift of a methane band relative to pure-methane as a proxy for the methane-nitrogen mixing ratio and band depth as a proxy for average depth into the surface. The average penetration depth of a photon at a particular wavelength is inversely related to the absorption coefficient at that wavelength, e.g. photons corresponding to larger absorption coefficients are absorbed more, preventing them from penetrating as deeply into the surface. In other words, stronger bands in the spectrum of an icy dwarf planet probe, on average, shallower into the surface than weaker bands.   \cite{lic06} found the weaker 0.72 $\mu$m methane band had a smaller shift than the stronger 0.89 $\mu$m band, suggesting a decreasing nitrogen content with increasing depth into Eris.  The authors speculated that as Eris moved toward aphelion and its atmosphere cooled, less volatile methane condensed out first. As the atmosphere cooled further, it become more and more nitrogen rich, until the much more volatile nitrogen could condense out on top of the methane-rich ice.  \cite{mer09} measured shifts for optical bands as well as the much stronger near-infrared bands that probe material close to the surface of Eris. They found very small shifts for the near-infrared bands, and so they suggested a nearly pure-methane layer, on top of a nitrogen-diluted layer, on top of a pure-methane layer for Eris.  

Our Eris analysis did not use shifts as a proxy for the methane-nitrogen mixing ratio. We found the methane-nitrogen mixing ratio that resulted in the best model fit to each observed band. Since each model band had absorption due to two methane phases and one of the phases, the highly-diluted methane phase, had a wavelength shift relative to the methane-rich phase that was band dependent \citep{qui97}, we found the model band profiles and wavelengths of maximum absorption were highly dependent on the the methane-nitrogen mixing ratio.  We found methane abundances of 8 $\pm$ 2\%, 9 $\pm$ 2\% and 10 $\pm$ 2\% for Eris'  0.73 $\mu$m, 0.87 $\mu$m, and 0.89 $\mu$m bands, i.e. we found no evidence of a trend in abundance with depth into the surface.  

Here we extend the work in \cite{teg10}. We describe the first laboratory experiment that shows the effect of crystallizing out a second methane phase on methane absorption bands. The experiment validates the importance of including two phases in modeling astronomical spectra of icy dwarf planets.  In addition, we  present three new near-infrared spectra of Pluto and a new optical spectrum of Triton. By combining the new spectra, previously published spectra, and our two phase model, we quantify the heterogeneity in the methane-nitrogen mixing ratio across the surface of Pluto and set limits on it across the surface of Triton. Furthermore, we find no evidence of a systematic trend in the methane-nitrogen mixing ratio with depth into the surfaces of Pluto, Triton, or Eris.  

\section{Laboratory Spectra}

The experiments reported here were done in a new laboratory ice facility located in the Department of Physics and
Astronomy of Northern Arizona University. Ice samples were crystallized in a cell fitted with windows to allow a spectrometer beam to pass through the ice (Figure 1). Detailed descriptions of the facility and our sample measurement procedures were published in \cite{teg10}. Subsequent to that paper, we made several important  improvements to the facility including the addition of a mercury cadmium telluride (MCT) type-A detector cooled with liquid nitrogen, off-axis aluminum paraboloid mirrors (Figure 2), and a 2 L mixing volume. Detailed descriptions of these improvements were published in \cite{gru11}.

The sample we describe here originated with high pressure cylinders of methane and nitrogen gases supplied by Airgas. The reported  purities of the gases  were $>$ 99.99 $\%$  for methane and $>$ 99.9 $\%$ for nitrogen. The cylinders were connected through separate pressure regulators and valves to the mixing volume. We mixed 10\% methane and  90\% nitrogen at room temperature in the mixing volume. Then,  we opened a valve which allowed the gas to flow into the empty, cold cell (T $=$ 65 K at the bottom of the cell), condensing it as a liquid. We froze the liquid by reducing the temperature in the cell at a rate of 0.1 K per minute to T $=$ 60 K at the bottom of the cell. During the freezing process, we maintaining a vertical thermal gradient of about 2 K across the sample by means of heaters (Figure 1). In this way, the sample froze from the bottom upward, with the location of the freezing front controlled by the cell temperature. Once the sample was frozen, we removed the vertical thermal gradient. 

Spectra were recorded with a Nicolet Nexus 670 Fourier transform infrared (FTIR) spectrometer at a sampling interval of 0.24 cm$^{-1}$, resulting in a spectral resolution of 0.6 cm$^{-1}$  (full width at half maximum of unresolved lines). We averaged over 100 spectral scans to improve the signal/noise ratio. After we recorded our initial ice spectrum, we ramped down to lower temperatures at  0.1 K per minute. We recorded spectra at temperatures of 60, 55, 50, 48, 46, 45, 43, 41, 40, 38, 36, 34, and 33 K at the bottom of the cell. Then, we ramped the temperature back up to 40 K at a rate of 0.5 K per minute, and took a last  spectrum of the ice sample. 

Raw transmission spectra were converted to absorption coefficient spectra as follows. First, we removed a low-amplitude, high-frequency interference pattern due to the windows in our cell by using a Fourier filter. Then, we removed water vapor lines in the spectra by dividing by a  background spectrum, i.e. the average of a spectrum taken just before insertion of the sample and another spectrum taken just after blowing off the sample).  Finally, we used the Beer-Lambert  law to compute  absorption coefficient spectra, $\alpha(\nu)$. 

Our spectra show that as we cooled the sample down we eventually exceeded the solubility limit of methane in nitrogen causing some methane molecules to migrate and crystalize out a second, methane-rich phase.  In Figure 3, we plot a portion of our absorption coefficient spectra showing the $\nu_2+\nu_3+\nu_4$ band at 1.72 $\mu$m for 60 K (red line), 50 K (reddish-purple line), 40 K (bluish-purple line), and 33 K (blue line). For reference, we plot pure-methane at 40 K as a black line. Clearly, the 60, 50, and 40 K spectra are blue shifted relative to the pure-methane spectrum due to the presence of nitrogen in the sample. The bands narrow and increase in height as the sample cools down.   At about 41 K, we began to see evidence of the second phase in our spectra. The onset of the second phase for a sample with 90\% nitrogen and T $\sim$ 41K is in agreement with the phase diagram of \cite{py83} which we show in Figure 4.  At our lowest temperature of  33 K in Figure 3, it is clear a second peak emerged toward the position of the pure-methane band. The second peak is due to the methane-rich phase. Upon increasing the temperature to 41 K, the second peak disappeared as did the methane rich crystals.  This is the first time anyone has spectroscopically observed such a phase transition in a methane-nitrogen ice.  It is clear that the phase transition has  a significant impact on the  band center and profile of methane ice bands.  

Crystallization of a second, methane-rich phase and not the $\beta$ $\rightarrow$  $\alpha$ transition of nitrogen was responsible for  the T $=$ 33 K band profile in Figure 3 (blue line).  We know this because  the 2.15 $\mu$m nitrogen band in the spectrum was dominated by a $\beta$-N$_2$ profile. Since the $\beta$ $\rightarrow$ $\alpha$ transition occurs at  T $=$ 35.6 K, we believe our temperature calibration was off by  2  to 3  K at our lowest temperature. Furthermore, we repeated the experiment described above with a sample  composed of 20\% methane and 80\% nitrogen because the phase diagram of \cite{py83} indicated the crystallization of the second, methane-rich phase should occur at a temperature  well above the 35.6 K nitrogen phase transition. Since the 20\% methane sample exhibited  the same behavior as the 10\% methane sample seen in Figure 3, there was no question the band shape at our lowest temperature in Figure 3 (blue line) was due to changes in methane and not nitrogen in the ice. 

\section{Astronomical Spectra}
Below we describe three new near-infrared spectra of Pluto taken with the NASA IRTF 3.0 m telescope, and a new optical spectrum of Triton taken with the  MMT 6.5 m telescope. In Table 1, we provide information about the circumstances of these new observations as well as our previously published optical spectrum of Pluto taken with the Bok telescope \citep{teg10} and near-infrared spectra of Triton taken with the IRTF \citep{gru10}. 

\subsection{Near-Infrared Spectra of Pluto}
We obtained near-infrared spectra of Pluto on the nights of 2007 Aug 8 UT, 2007 Aug 10 UT, and 2010 Jul 1 UT using the NASA IRTF  3.0 m telescope and the SpeX spectrograph as part of a long term project to monitor the spatial distribution of ices on Pluto. We used SpeX in short cross-dispersed mode with a 0.5 arc sec x 15 arc sec slit  that provided wavelength coverage of 0.8 to 2.4 $\mu$m at a resolution ($\lambda$/fwhm) of 1300-1400 and a sampling of 0.001 $\mu$m pixel$^{-1}$.  There were some thin clouds and the seeing was $\sim$ 0.6 arc sec on all three nights. Observations began at an airmass of about 1.4,  transited the meridian at at airmass of about 1.2, and ended at an airmass of about 1.4 on all three nights.  Spectra of Pluto were bracketed by spectra of an argon arc lamp for wavelength calibration.  The arc calibration was checked against telluric sky emission lines extracted from Pluto frames, which sometimes called for a slight shift relative to the argon spectra, due to flexure of SpeX.  Spectra of the solar analog star HD 160788 were taken on 2007 Aug 8 UT and 2007 Aug 10 UT in order to remove instrumental signatures, telluric bands, Fraunhofer lines, and the solar continuum. Spectra of the solar analog star HD 170379 were taken on 2010 Jul 01 UT for the same purpose. The summed Pluto spectra of Aug 2007 8 UT, 2007 Aug 10 UT,  and 2010 July 1 UT had total exposure times of 108, 116, and 112 minutes.

In Figure 5, we plot our relative reflectance spectrum of Pluto (bottom) for 2007 Aug 10 UT and Triton (top) for 2007 Jun 22 UT
\citep{gru10}. Our Triton spectrum was shifted vertically upward on the plot by 0.4 to facilitate a comparison between the two spectra. The wavelengths of pure-methane ice absorption bands are marked by ticks \citep{gru02}.  Bands at 1.97 $\mu$m, 2.01 $\mu$m, and 2.07 $\mu$m in Triton's spectrum are due to carbon dioxide ice. The band at 2.15 $\mu$m  in both spectra is due to N$_2$ ice.  Clearly, Pluto has deeper methane bands than Triton. 

The 2007 Aug 08 UT and 2010 Jul 01 UT spectra of Pluto have some important differences from the 2007 Aug 10 UT spectrum of Pluto shown in Figure 5; however, they are difficult to see on the scale of Figure 5. The same can be said of the 2007 Jun 25 UT spectrum of Triton and the 2007 Jun 22 UT spectrum of Triton shown in Figure 5. We will describe the important differences in the pages that follow.  The purpose of Figure 5 is to describe the quality of our IRTF spectra and compare the near-infrared spectra of Pluto and Triton. 

\subsection{Optical Spectra of Triton}
We obtained optical spectra of Triton on the night of 2010 July 13 UT using the MMT 6.5 m Telescope, Red Channel Spectrograph, and a red sensitive, deep depletion CCD. We used a 1'' x 180'' entrance slit and a 270 grooves mm$^{-1}$ grating that provided wavelength coverage of 0.64 to 0.99 $\mu$m  and a dispersion of 0.0004 $\mu$m  pixel$^{-1}$ in first order. There were passing clouds and the seeing was $\sim$ 1.5 arc sec. Triton was placed at the center of the slit and the telescope was tracked at the sidereal rate in order to use the autoguider. We centered Triton in the slit every six to ten minutes.  At the start of our observations, Triton was on the meridian and had an airmass of 1.4, and at the end of our observations it had an airmass of 1.5.  Triton  spectra were bracketed by HeNeAr spectra in order to obtain an accurate wavelength calibration. Spectra of the solar analog star  HD 202282 were taken to remove telluric bands, Fraunhofer lines, and the solar continuum.  The airmass differences between the Triton spectra and the HD 202282 spectra were $\sim$ 0.1 airmass. We used standard procedures to calibrate and extract one-dimensional spectra from two-dimensional spectral images \citep{mas92}. 

Since we saw no significant difference between the 24 120-s exposures of Triton, we summed the individual exposures to yield a single exposure with a 48 minute exposure time. To assess the uncertainty in our wavelength calibration, we measured the wavelengths of well-resolved sky emission lines in our Triton spectrum and compared them to the VLT high spectral resolution sky line atlas of \cite{han03}. The standard deviation of the differences between our centroids and Hanuschik's centroids was $\sim$ 0.00004 $\mu$m, suggesting our wavelength calibration was good to $\sim$0.00004 $\mu$m or $\sim$ 0.1 pixel. 

In Figure 6, we present the resulting reflectance spectrum of Triton (top) and compare it to spectra of  Pluto (middle) and Eris (bottom) from \cite{teg10}. All three spectra were normalized to 1.0 between 0.82 and 0.84 $\mu$m. The Pluto and Triton spectra were shifted upward on the plot by 0.2 and 0.4 to facilitate comparison of the spectra.  The wavelengths of pure-methane ice absorption bands are marked by ticks \citep{gru02}. Despite a continuum signal-to-noise ratio reaching $\sim$ 400, the Triton spectrum exhibits only one obvious methane band at $\sim$0.89 $\mu$m. Clearly,  Eris has the most optical methane absorption, followed by Pluto, and then Triton. 

\section{Model}
We used Hapke theory \citep{hap93} to transform laboratory optical constants into spectra suitable for comparison to Pluto, Triton, and Eris
reflectance spectra. Hapke theory accounts for multiple scattering of light within a surface composed of particulate ice. We used Hapke model parameters of h $=$ 0.1, B$_o$ $=$ 0.8,  $\overline{\theta}$ $=$ 30$^{\circ}$ , P(g) $=$ a two component Henyey-Greenstein function with 80\% in the forward scattering lobe and 20\% in the back scattering lobe, both lobes having asymmetry parameter a $=$ 0.63. It is important to recognize that these Hapke parameters are not unique; however, they are plausible values for scattering in methane-nitrogen ice and are comparable to Hapke parameters in previous fits of icy dwarf spectra, see e.g. \cite{gru01} and \cite{teg10}. Our implementation of the Hapke model used two particle sizes $D_1$ and $D_2$ to account for the observed enhancement of weak methane bands relative to stronger methane bands. "Particle sizes" $D_1$ and $D_2$ should be thought of as more generally representing typical
distances traversed within the ice between scattering events, and are not necessarily sizes of discrete particles, since the texture of the ice is uncertain. See \cite{elu07} and references therein. We also added a tholin-like absorber to account for the reddish slope in the optical part of the spectrum.

The binary phase diagram of \cite{py83} and our lab experiments reported here show that it is essential to include two methane ice components in our Hapke model $-$ methane-rich crystals near the nitrogen solubility limit and nitrogen-rich crystals near the methane solubility limit. The best approach is to use laboratory absorption coefficient spectra of ice samples exhibiting the two components. Unfortunately, we have only begun lab work to map out methane and nitrogen absorption coefficient spectra at a suite of compositions and temperatures relevant to icy dwarf planets. On the other hand, it is possible to approximate the two ice components using previous work on pure-methane \citep{gru02}  and highly-diluted methane \citep{qui97} ice samples. Specifically, we took $f$ as the overall methane abundance, $\eta$ as the fraction of methane rich crystals, $S_{N_2}$ as the solubility limit of nitrogen in methane, and $S_{CH_4}$  as the solubility limit of methane in nitrogen. From the methane$-$nitrogen phase diagram of \cite{py83}, we estimated $S_{N_2}$ $=$ 3\% and $S_{CH_4}$ $=$ 5\% at T = 40 K for Pluto and Triton, and $S_{N_2}$ $=$ 3\% and $S_{CH_4}$ $=$ 2\% at T = 30 K for Eris.  In \cite{teg10} , we did not take the temperature variation of the solubilities into account and used $S_{N_2}$ $=$ 4\% and $S_{CH_4}$ $=$ 2\% for both Pluto and Eris.  Our analysis here recomputes the Pluto and Eris  methane abundances in \cite{teg10} taking the temperature dependence of the solubilities into account. The relationship between $f$, $\eta$, and the solubilites was given by

\begin{eqnarray}
f &=& \eta(1-S_{N_2}) + (1-\eta)S_{CH_4}.
\end{eqnarray}

\noindent Solving for $\eta$, we got

\begin{eqnarray}
\eta & =&{{f-S_{CH_4}}\over{1-S_{CH_4}-S_{N_2}}} .
\end{eqnarray}

\noindent  We combined absorption coefficients of the methane-rich crystals ($\alpha_u$), and absorption coefficients of the highly-diluted methane crystals  ($\alpha_s$) as follows, 

\begin{eqnarray}
\alpha(\nu) &=&c[ \eta(1-S_{N_2})\alpha_u(\nu) + (1-\eta)S_{CH_4}\alpha_s(\nu)]
\end{eqnarray}

\noindent where

\begin{eqnarray}
c&=&{[\eta(1-S_{N_2})+(1-\eta)S_{CH_4}]^{-1}}
\end{eqnarray}

\noindent We approximated $\alpha_u$ with  pure-methane  coefficients, and $\alpha_s$ with  pure-methane coefficients blue shifted by amounts observed in highly-diluted samples  \citep{qui97}. It is important to recognize that the blue shift depends on the band. In Table 2, we give the wavelengths in microns (column 2) and frequencies in wavenumbers (column 4) of six pure-methane ice bands and the blue shifts of the bands in highly-diluted samples in microns (column 3) and wave numbers (column 5). Notice the shifts differ by a factor of two in wavelength and a factor of three in frequency. 

We explored parameters specific to our implementation of the Hapke model ($D_1$ and $D_2$,  percent of large scatters by volume, tholin-like absorption, and bulk methane abundance) with Monte Carlo techniques. For each methane ice band in spectra of Pluto, Triton, and Eris, we ran $\sim$ 50,000 models and found the set of parameters that gave the best $\chi^2$ goodness of fit to the observed band. 

In the middle  of Figure 7, we plot the 1.72 $\mu$m methane band in the 2007 Aug 10 UT spectrum of Pluto (black line) and the best fit Hapke model that includes the two methane phases and an overall bulk abundance of 8.1\% methane and 91.9\% nitrogen (red line). In the best fit model, about 60\% of the methane is in the nitrogen-rich crystals and the remainder is in the methane-rich crystals. The best fit model does a very good job matching the wavelength of maximum absorption and profile of the Pluto band. For reference, we plot a Hapke model of pure-methane (blue line) and a Hapke model of highly-diluted methane (cyan line).  Clearly, the individual phases do not match Pluto's band.  The band profiles of the individual phases are too narrow compared to the Pluto band. In addition, the methane-rich and nitrogen-rich phases have wavelengths of maximum absorption that are red shifted and blue shifted relative to Pluto's wavelength of maximum absorption.  

In the upper part of Figure 7, we plot the 1.72 $\mu$m methane band in the 2007 Jun 22 UT spectrum of Triton (black line) and the best fit Hapke model that includes the two methane phases and an overall bulk abundance of 5.5\% methane and 94.5\% nitrogen (red line). In the best fit model, about 90\% of the methane is  in the nitrogen-rich crystals and the remainder is in the methane-rich crystals.  It is clear that the Triton band is much narrower and blue shifted relative to the Pluto band.  We find the highly-diluted model is almost as good a fit to the Triton band as the two phase model.

In the lower part of Figure 7, we plot the 1.72 $\mu$m methane band in a  spectrum of Eris \citep{dum07} (black line) and the best fit Hapke model that includes the two methane phases and an overall bulk abundance of 8\% methane and 92\% nitrogen (red line). In the best fit model, about 23\% of the methane is in the nitrogen-rich crystals and the remainder is  in the methane-rich crystals.  It is clear the Eris band is closest to the position of the pure methane band. 

With the exception of the 1.67 $\mu$m  methane band which is contaminated by the 1.5$\mu$m water band in Triton's spectrum
\citep{gy04}, we fit models to the five strongest methane bands in our IRTF spectra of Pluto and Triton. We did not fit  models to bands with wavelengths $<$ 1.1 $\mu$m because we did not have a sufficient number of airglow lines to remove wavelength shifts caused by flexure in the IRTF spectra.  Our optical MMT and Bok spectra had excellent wavelength calibrations for the 0.89 $\mu$m methane bands due to frequent observation of HeNeAr spectra, so we fit models to these bands. The \cite{dum07} spectrum of Eris contained three of the five bands we fit in our Pluto and Triton spectra, so we fit these bands as well.  In Tables 3, 4 and 5 we give methane abundances corresponding to the best models.

\section{Results}

\subsection{Surface Heterogeneity}

We used our abundances in Tables  3  and 4 to look for  differences in the methane-nitrogen stoichiometry across the surfaces of Pluto and Triton. From Table 1 and the column headings of Table 3, we see that the sub-Earth longitudes of our three IRTF spectra of Pluto were 10$^{\circ}$, 125$^{\circ}$, and 257$^{\circ}$.  From Tables 1 and 4, we see that the sub-Earth longitudes of our two IRTF spectra of Triton were 138$^{\circ}$ and 314$^{\circ}$.

The values in Table 3 suggest heterogeneity in the methane abundance across the surface of Pluto. First, notice that each abundance at a sub-Earth longitude of 257$^{\circ}$ (column 5) is larger that the corresponding abundance at a sub-Earth longitude of 125$^{\circ}$ (column 3). Furthermore, each abundance at a longitude of 10$^{\circ}$ (column 2) is even larger than the corresponding abundance  at a longitude of 257$^{\circ}$ (column 5).  We quantified the significance of this apparent pattern by  applying  the non-parametric Wilcoxon rank sum test to the values in Table 3.  We found a 0.8\% probability that the abundances at sub-Earth longitudes of 10$^{\circ}$ and 125$^{\circ}$ have the same mean. In other words, the difference is statistically significant. 

As for the Triton abundances in Table 4, notice that each of the  abundances at a sub-Earth longitude of  314$^{\circ}$ (column 3) is equal to or larger than the corresponding abundance at a sub-Earth longitude of 138$^{\circ}$. We used the Wilcoxon rank sum test and found there is a 29\% probability that the abundances from the two longitudes have the same mean. In other words, the difference is not statistically significant. It is important to recognize that our observations and model are not able to measure methane abundances smaller than $S_{CH4}$, i.e. 5\% for Triton.

In Table 5, we present methane abundances for Eris taken by \cite{teg10} on 2008 October 3 UT as well as our analysis of a spectrum taken by \cite{dum07} on 2005 Oct 18 UT. As there is only a tentative detection of Eris' light curve and rotation period \citep{roe08}, we cannot relate the spectra to sub-Earth longitudes and so we cannot make any statements about Eris' surface heterogeneity other than two random longitude samples gave similar methane abundances.  

\subsection{Stratigraphy}

Next, we looked for changes in the methane-nitrogen mixing ratio as a function of depth into the surfaces of Pluto and Triton. In Figure 8, we plot the  percentage of methane in Table 3 vs. the relative reflectance at band minimum for Pluto.  Stronger bands that probe on average shallower depths into the surface are located toward the left side of the figure, and weaker bands that probe deeper into the surface are located toward the right side of the figure.  The cyan squares come from our spectrum taken at a sub-Earth longitude of 10$^{\circ}$,  black circles come from our observations at a sub-Earth longitude of 257$^{\circ}$, and the red squares come from our observations at  sub-Earth longitude of 125$^{\circ}$.  The horizontal lines correspond to average methane abundances of 9.1\%, 8.2\% and 7.1\% (see Table 3).  The error bars have lengths of $\pm$ 1$\sigma$ (see Table 3).  It is clear from the figure that there is no apparent trend between the mixing ratio and depth for any of the three longitudes. 

In Figure 9, we plot percentage of methane in Table 4 vs. the relative reflectance at band minimum for Triton. Again, stronger bands that probe on average shallower depths into the surface are located toward the left side of the figure, and weaker bands that probe deeper into the surface are located toward the right side of the figure.  The black circles come from our spectrum taken at a sub-Earth longitude of 314$^{\circ}$, and the red squares come from our observations at a sub-Earth longitude of 138$^{\circ}$. The horizontal lines corresponds to average methane abundances of 5.3\% and 5.0\% (see Table 4).  The error bars have lengths of $\pm$ 1$\sigma$ (see Table 4).  It is clear from the figure that there is no apparent trend between the mixing ratio and depth for either longitude.

As for Eris, we are limited by the fact we do not know the sub-Earth longitudes corresponding to the optical \citep{teg10} and near-infrared \citep{dum07} spectra. If we assume homogeneity across the surface or a nearly pole-on viewing geometry, then Table 5 suggests Eris  too exhibits no evidence of a change in the mixing ratio with depth.  

To get an idea of the depths probed by these observations, we ran Monte Carlo ray-tracing models \citep{gru00} for the D, $\omega$, and P(g) configurations in our Hapke models.  The mean depths sampled for normally incident photons escaping from the surface are shown versus the fraction escaping (hemispheric albedo A$_H$) in Figure 10.  A$_H$ is not the same quantity as the disk integrated albedos observed at the telescope, but we use it as a proxy because the latter are difficult to estimate via multiple scattering ray-tracing techniques.  In this computation, we assumed a somewhat arbitrary filling factor of 50\%.  A higher filling factor would imply shallower sampling, such that photons traverse the same quantity of ice before escaping.  The single scattering phase function P(g) has a strong effect on depths probed, but it is coupled to the spatial scale of the scattering represented by D in such a way as to lessen the sensitivity of the mean depth probed to the assumed P(g).  A more forward scattering P(g) requires a larger average number of scatters to scatter initially downward trending photons back out of the surface such that they can be observed, but the greater number of scattering events implies smaller D in order for the same quantity of ice to be traversed so that the absorption bands produced are of the observed strengths. It appears our spectra are sampling depths on the scale of centimeters. 

\section{Discussion}
 
We found our largest methane abundance on Pluto's surface at a longitude of 10$^{\circ}$. For reference, the  sub-Charon hemisphere is centered at 0$^{\circ}$ longitude. \cite{gru01} found the smallest shifts in  methane bands, i.e. the largest methane abundance, near a sub-Earth longitude of 0$^{\circ}$.   We found our smallest  methane abundance at 125$^{\circ}$. \cite{gru01} found the largest shifts in the bands, i.e. the smallest methane abundance, near this longitude (see their Figure 9). In short, our quantitative measurements of methane abundance are in good agreement  with Grundy and Buie's qualitative results derived from band shifts. It is interesting to note that the smallest methane abundance occurs at longitudes where the  nitrogen optical path lengths are the longest \citep{gru01}.  

It is even more interesting to note the small amplitude of the heterogeneity across Pluto's surface. \cite{gru01} showed the greatest difference in shifts of methane bands, and so the greatest contrast between methane-nitrogen mixing ratios, occurred between longitudes near 10$^{\circ}$ and 125$^{\circ}$.  We found the methane-nitrogen mixing ratio at these longitudes differs by only $\sim$2\%. In the middle of Figure 11, we plot the 1.72 $\mu$m band from our spectra at longitudes of 10$^{\circ}$ (black line) and 125$^{\circ}$ (red line) to illustrate the small but measurable effect the change in methane abundance has on the band.  The two vertical dashed lines indicate the location of maximum absorption for highly-diluted methane (1.7198 $\mu$m) and pure-methane (1.7245 $\mu$m). Clearly, the band from a longitude of 10$^{\circ}$ (black line) is slightly broader on the redward side of the profile than the band from a longitude of 125$^{\circ}$ (red line). The broader nature is due to the presence of a larger amount of the methane-rich phase.

As for Triton, we  found methane abundances of 5.0 $\pm$ 0.1\% and 5.3 $\pm$ 0.4\% for sub-Earth longitudes of 138$^{\circ}$ and 314$^{\circ}$.  We point out that \cite{gru10} found the largest shifts in methane bands near a longitude $\sim$ 90$^{\circ}$ and the smallest shifts in methane bands  near a longitude  $\sim$ 270$^{\circ}$ (see their Figure 6).  Again, our quantitative abundances are in good agreement with the Grundy et al. qualitative abundances. In the upper part of Figure 11, we plot the 1.72 $\mu$m bands in our Triton spectra for sub-Earth longitudes of 314$^{\circ}$ (black line) and 138$^{\circ}$ (red line). Although both bands are narrower and blue shifted relative to the Pluto bands in the middle part of the figure, there is no obvious difference between the two Triton band profiles. But then, we find the mixing ratio for the two Triton spectra differ by only 0.3\%. 

Pluto, Triton, and Eris do not exhibit  any evidence of a trend between methane-nitrogen mixing ratio and depth below the surface (Figures 8 and 9 and Table 5). This lack of a trend is counter to the expectations of \cite{gru00}, who noted that visible sunlight is able to penetrate deeper into a nitrogen-rich surface than the depths from which thermal emission emerges.  This mis-match should preferentially drive more volatile nitrogen from the depths where sunlight is absorbed to the shallower depths affected by radiative cooling.  One possible reason that we did not see the expected enhancement of methane abundance at depth could be that photons reflected at wavelengths with diagnostic $CH_4$ absorption bands do not probe deep enough to reach the depths where sunlight is absorbed on Pluto, Triton, and Eris, and thus we do not see the effect.  Another possibility is that the ice is so densely sintered (e.g., \cite{elu07}) that nitrogen gas is unable to flow through it, as required by the \cite{gru00} model.

\section{Future Work}
A proper analysis of methane bands in icy dwarf spectra must take the nearly-pure-methane crystals and highly-diluted methane crystals into account. Here we approximated the nearly-pure crystals by using pure-methane absorption coefficients and the highly-diluted phase by shifting pure-methane coefficients by amounts seen for highly diluted samples. The best way to take the two phases into account is to use absorption coefficients of samples that include the two phases. We have coefficients for two such samples, methane and nitrogen abundances of 10\% and 90\% as well as 20\% and 80\%. We are in the midst of generating coefficients for samples at a suite of temperatures and concentrations relevant to icy dwarf planet spectra.  Such coefficients will yield the best methane-nitrogen mixing ratios for icy dwarf planets. 

The lack of an increase in the methane abundance with an increase in depth into the surface of Pluto was a surprise to us. In the future, we will probe deeper into the surface of  Pluto by observing its 0.73 $\mu$m and 0.89$\mu$m bands at sub-Earth longitudes of 10$^{\circ}$, 125$^{\circ}$, and 257$^{\circ}$ (see Table 3 and  Figure 8).  

\acknowledgments
We are grateful to the NASA Planetary Astronomy Program (NNX10AB24G) for support of the telescope observations and their analysis and the  NASA Outer Planets Research Program (NNX11AM53G)  for support of the laboratory part of this project. 
We thank the NSF REU Program and the NASA Spacegrant Program at Northern Arizona University for supporting student work on this project. We are grateful to the NASA IRTF and Steward Observatory TACs for consistent allocation of telescope time. We thank C. Dumas for a digital copy of his Eris spectrum.

\clearpage

\begin{figure}
\epsscale{.80}
\plotone{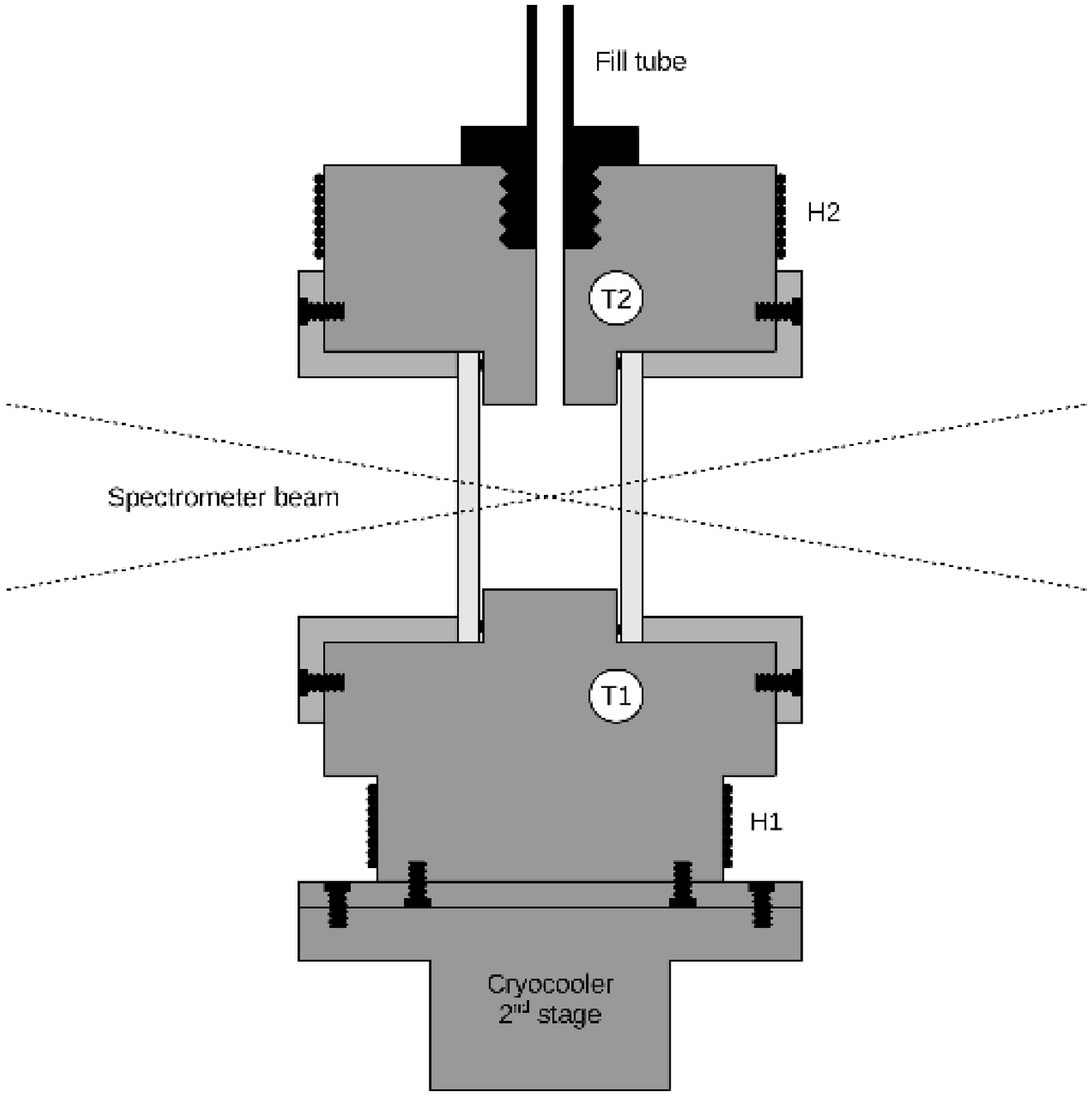}
\caption{Schematic diagram of the cell in cross section. The cell is mounted on top of the cryocooler.  Gas enters the cell from above via a fill tube. The dotted lines represent the spectrometer beam through the sample. Thermometers T1 and T2 and heating elements H1 and H2 control the temperature of the sample  as well as a vertical temperature gradient across the sample. Further details concerning the cell are described in \cite{teg10} and \cite{gru11}.}
\end{figure}

\clearpage

\begin{figure}
\epsscale{.80}
\plotone{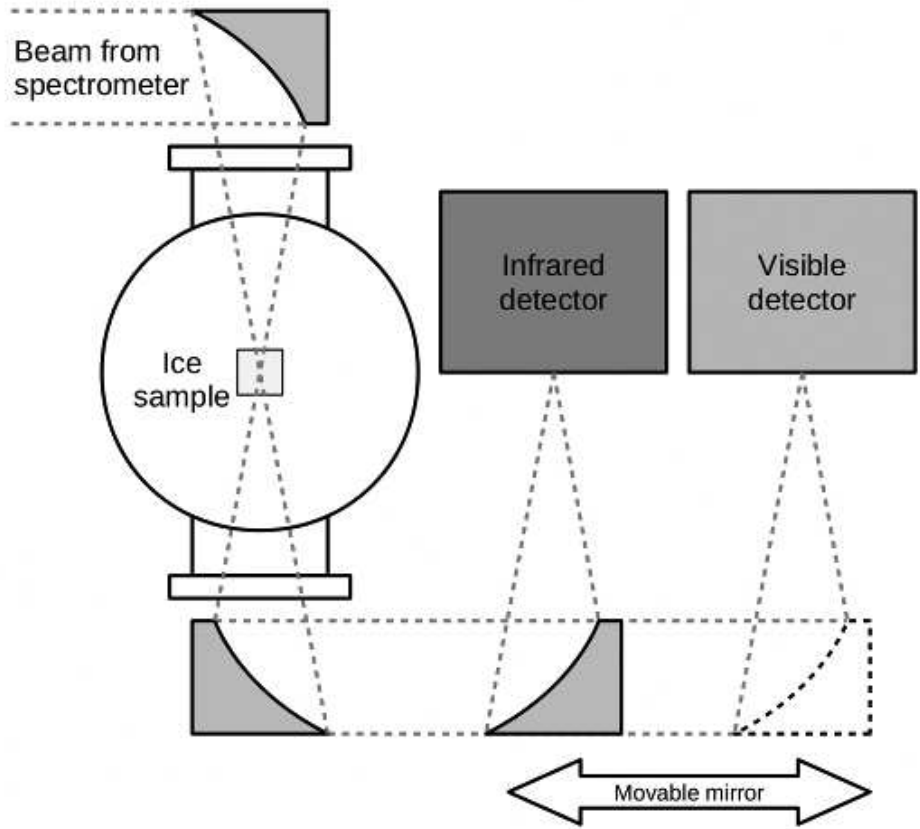}
\caption{The optical train in our experiment. The spectrometer beam is represented by dashed lines. The first  off-axis paraboloid mirror focuses the beam into the sample volume, then two more identical mirrors re-focus the beam onto one of two detectors, an infrared or visible detector. The third mirror in the optical train is movable and enables us to select the infrared or visible detector. Only the infrared detector was used in the experiment described here.}
\end{figure}

\clearpage

\begin{figure}
\epsscale{.80}
\plotone{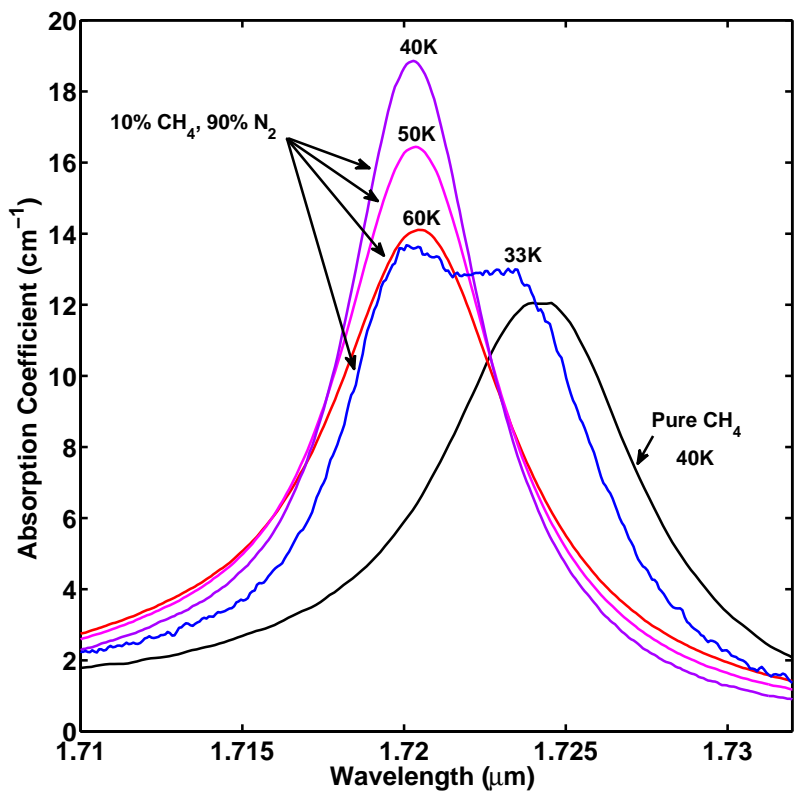}
\caption{A portion of our absorption coefficient spectra isolating the $\nu_2+\nu_3+\nu_4$ methane ice band  in our 10\% methane and 90\% nitrogen sample at 60 K (red line), 50 K (reddish-purple line), and 40 K (blueish-purple line). For reference, we plot the same band for a pure-methane sample at 40 K (black line). The presence of nitrogen blue shifted the bands relative to the pure-methane band. In addition, the bands narrowed and increased in height during the cooling from 60 K to 50 K to 40 K. Upon further cooling, the band began to broaden and then grew a second peak as seen in the sample at 33 K (blue line).  The second peak is due to our exceeding the solubility limit of methane in nitrogen, resulting in migration of methane molecules, and the  formation of a second phase consisting of crystals rich in methane. Upon warming the sample back up to 40 K, the second peak disappeared from the band profile as did the methane rich crystals.}
\end{figure}

\clearpage

\begin{figure}
\epsscale{.70}
\plotone{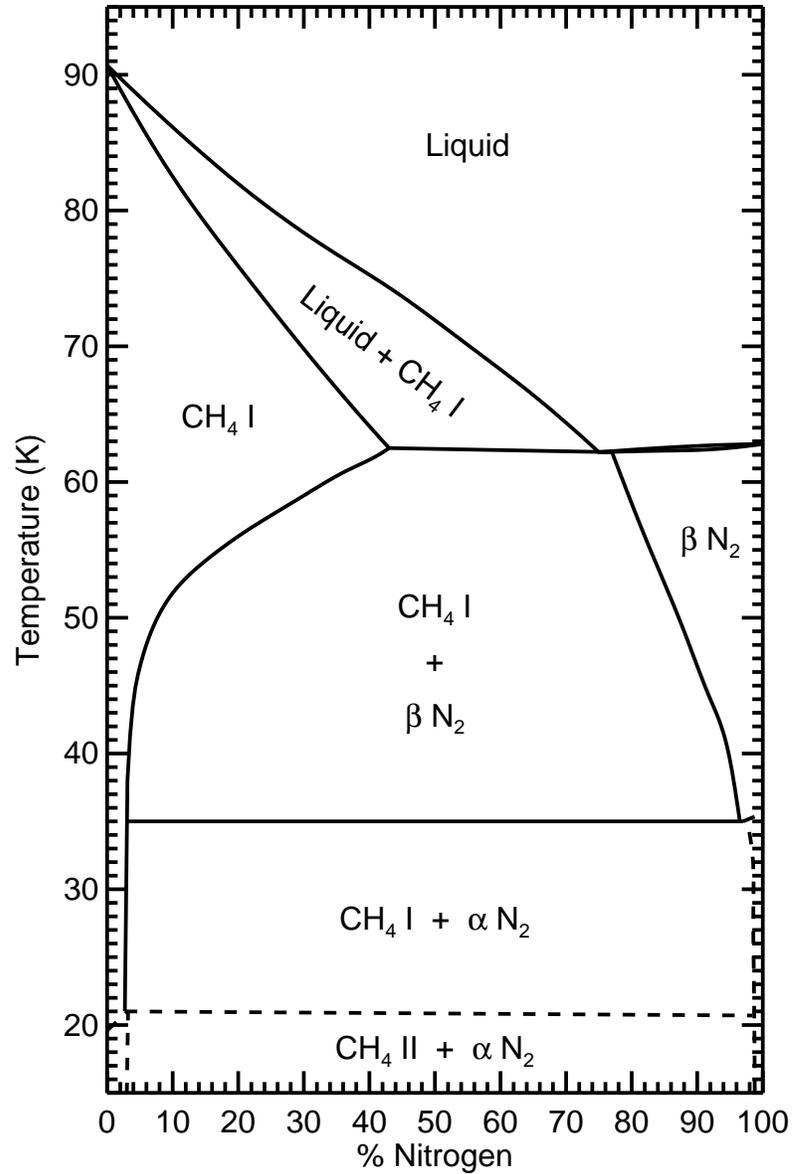}
\caption{Binary phase diagram for methane-nitrogen ice mixtures from \cite{py83}. In our experiment with an ice sample consisting of  90\% nitrogen, we witnessed the growth of a second phase consisting of methane-rich crystals in our spectra at about 41K,  in agreement with the phase diagram. }
\end{figure}

\clearpage

\begin{figure}
\epsscale{.80}
\plotone{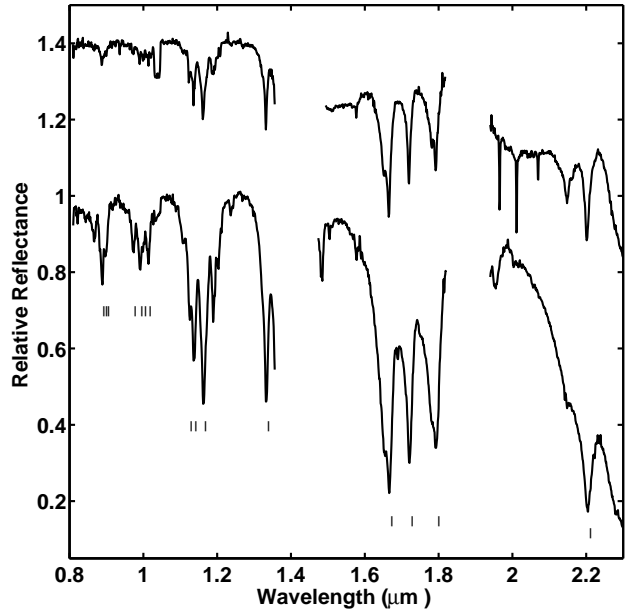}
\caption{Near-infrared relative reflectance spectra of Pluto (bottom) taken with the NASA IRTF  on 2007 Aug 10 UT compared to a spectrum of  Triton (top). The Triton spectrum is from  \cite{gru10}. The Triton spectrum was shifted vertically upward by 0.4 to facilitate a comparison between the two spectra. Pure-methane ice absorption bands are marked by ticks \citep{gru02}. Pluto has stronger near-infrared methane absorption bands than Triton.}
\end{figure}

\begin{figure}
\epsscale{.80}
\plotone{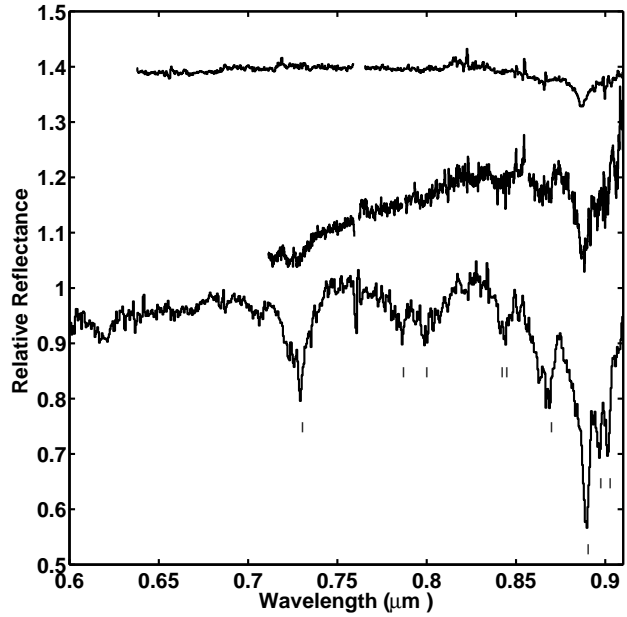}
\caption{Optical relative reflectance spectrum of Triton (top) taken with the MMT 6.5 m telescope on 2010 July 13 UT compared to spectra of Pluto (middle) and Eris (bottom). The Pluto and Eris spectra are from \cite{teg10}. All spectra were normalized to 1.0 over the 0.82 to 0.84 $\mu$m wavelength range. The Pluto and Triton spectra were moved upward by 0.2 and 0.4 to facilitate a comparison between the three spectra. Pure-methane ice absorption bands are marked by ticks \citep{gru02}. Eris has the strongest optical methane absorption bands, followed by Pluto, and then Triton.}
\end{figure}

\clearpage

\begin{figure}
\epsscale{.80}
\plotone{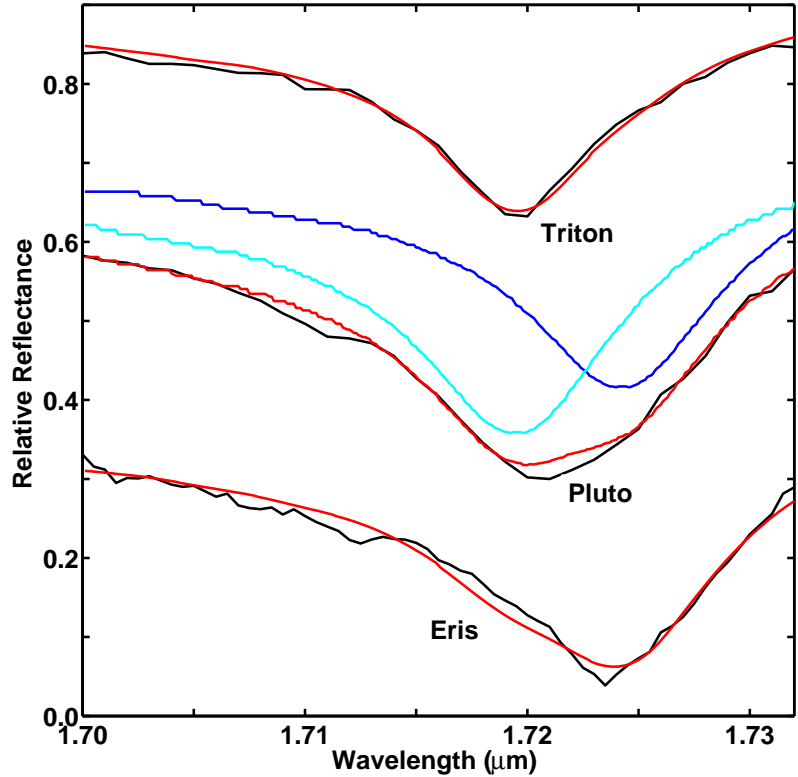}
\caption{The middle of the plot  shows  a portion of the IRTF spectrum of Pluto taken on 2007 Aug 10 UT  isolating the 1.72 $\mu$m methane ice band (black line), the best fit Hapke model including methane-rich and nitrogen-rich crystals and a bulk methane and nitrogen abundance of 8.1\% and 91.9\% (red line). For comparison, we include Hapke models of the individual phases approximated by pure-methane (blue line) and highly-diluted methane (cyan line). The two phase model does a very good job matching the wavelength of maximum absorption and profile of Pluto's band. The band profiles of the individual phases are far too narrow compared to Pluto's band. The upper part of the plot shows a portion of the IRTF spectrum of Trition taken on 2007 Jun 22 UT (black line), and the best fit Hapke model with bulk methane and nitrogen abundance of 5.5\% and 94.5\% (red line).  Triton's band closely matches the highly-diluted band (cyan line). The lower part of the plot shows a portion of Eris' spectrum \citep{dum07} and the best fit Hapke model with bulk methane and nitrogen abundances of 8\% and 92\% (red line). The Eris band is closest to the position of the pure methane band (blue line).}
\end{figure}

\begin{figure}
\epsscale{.80}
\plotone{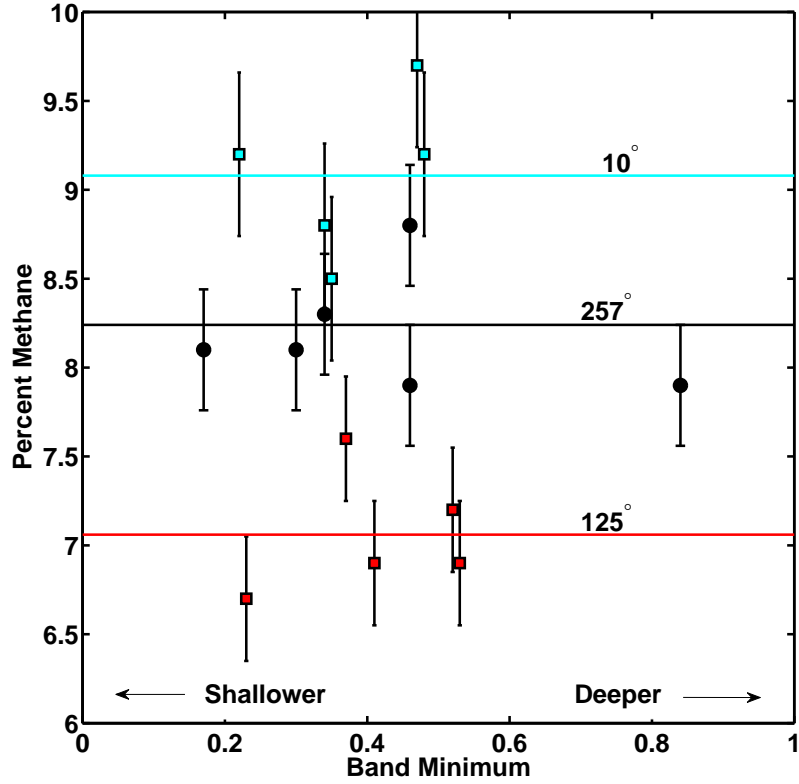}
\caption{Methane mixing ratio as a function of relative reflectance at band minimum for observations of Pluto on 2007 Aug 8 UT at a longitude of 10$^{\circ}$ (cyan),  2007 Aug 10 UT at a longitude of 257$^{\circ}$ (black),  and 2010 Jul 1 UT at a longitude of 125$^{\circ}$ (red).  Stronger bands (smaller band minima) probe on average shallower into the surface than weaker bands (larger band minima).  Neither longitude shows any evidence of a trend between mixing ratio and  band minima. There is no evidence of a compositional gradient with depth into the surface of Pluto in our observations.} 
\end{figure}

\begin{figure}
\epsscale{.80}
\plotone{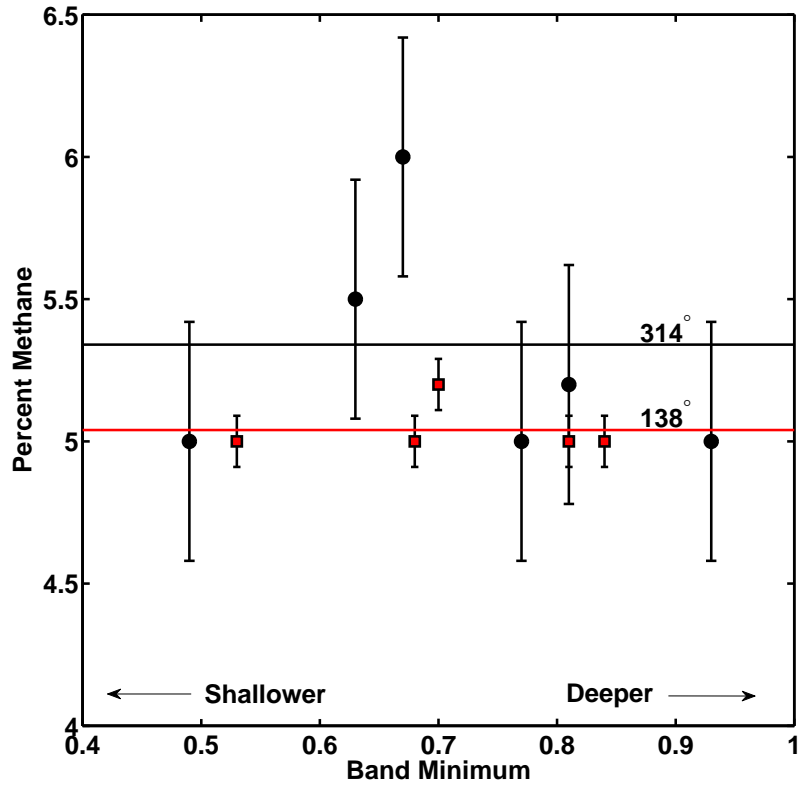}
\caption{Methane mixing ratio as a function of relative reflectance at band minimum for observations of Triton on 2007 Jun 22 UT at a longitude of 314$^{\circ}$ (black) and 2007 Jun 25 UT at a longitude of 138$^{\circ}$ (red).  Stronger bands (smaller band minima) probe on average shallower into the surface than weaker bands (larger band minima).  Neither longitude shows any evidence of a trend between mixing ratio and  band minima. There is no evidence of a compositional gradient with depth into the surface of Triton in our observations.} 
\end{figure}

\begin{figure}
\epsscale{.80}
\plotone{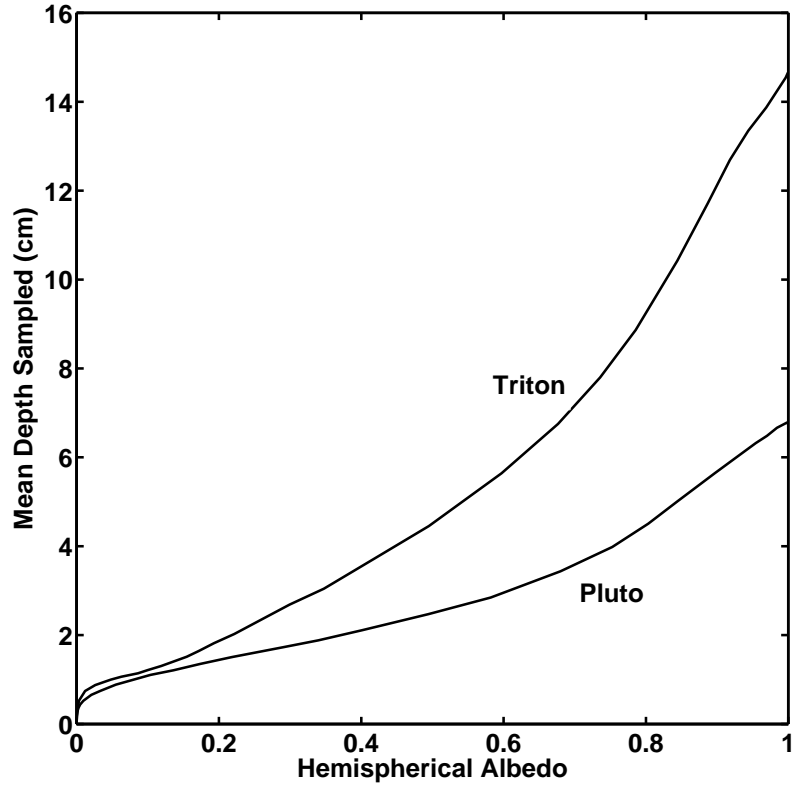}
\caption{Relation between mean depth probed by photons normally incident on a surface described by our Hapke models and hemispherical albedo as revealed by a  Monte Carlo ray tracing model.  Because our single scattering phase function P(g) is forward scattering, the mean depth sampled tends to be substantially larger than D, especially for higher albedos where photons are, on average, scattered many times before they are absorbed or escape from the surface.}
\end{figure}

\begin{figure}
\epsscale{.80}
\plotone{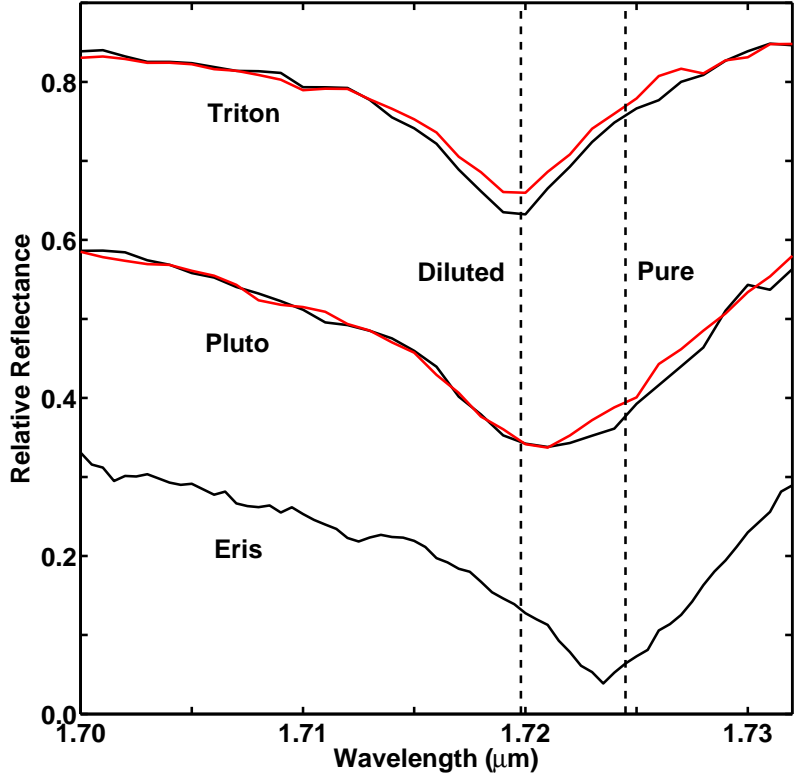}
\caption{In the middle of  the plot, we show the 1.72 $\mu$m band in the 2007 Aug 8 UT spectrum of Pluto corresponding to a sub-Earth longitude of 10$^{\circ}$ (black line) and the 2010 Jul 01 UT spectrum of Pluto corresponding to a sub-Earth longitude of 125$^{\circ}$ (red line). The slightly larger methane abundance at a longitude of 10$^{\circ}$ is apparent from the slightly broader nature of its redward profile compared to the band at a longitude of 125$^{\circ}$. The vertical lines at 1.7198 $\mu$m and 1.7245 $\mu$m are the locations of maximum absorption for highly-diluted and pure-methane. In the upper part of the plot, we show the 1.72 $\mu$m bands in Triton's spectra for 2007 Jun 22 UT corresponding to sub-Earth longitude of 314$^{\circ}$ (black line) and 2007 June 25 UT corresponding to a sub-Earth longitude of 138$^{\circ}$ (red line).  Although both Triton bands are narrower and blue shifted compared to the Pluto bands due to most of the methane being  in nitrogen-rich crystals, there is no measurable difference between the two Triton band profiles. In the lower part of the plot, we show the 1.72 $\mu$m band in Eris' spectrum \citep{dum07}. Its minimum is closest to pure methane because it has the largest amount of methane in methane-rich crystals of the three icy dwarf planets. } 
\end{figure}

\clearpage

\begin{deluxetable}{cccccccc}
\tablewidth{0pt}
\tablecaption{Circumstances of Observations}
\tablehead{
\colhead{ Object} & 
\colhead{Telescope} &
\colhead{UT Date} & 
\colhead{UT Time} &
\colhead{Lon$^a$} &
\colhead{Lat$^a$} &
\colhead{Exp Time} &
\colhead{Ref}
\\
\colhead{} & 
\colhead{} & 
\colhead{} & 
\colhead{} &
\colhead{} & 
\colhead{} &
\colhead{(min)} &
\colhead{}}
\startdata
Pluto  & Bok   & 2007 Jun 21 &  07:13 & 194 & $+$40 &120  & b \\
Pluto  & IRTF & 2007 Aug 08 &  07:04 & 010 & $+$39 & 108 & c\\
Pluto  & IRTF & 2007 Aug 10 &  07:04 & 257 & $+$39 & 116 & c\\
Pluto  & IRTF & 2010 Jul  01 & 10:00  & 125 & $+$44 & 112 & c\\
Triton & IRTF & 2007 Jun 22  & 13:10  & 314 & $-$48 & 160 & d\\
Triton & IRTF & 2007 Jun 25  & 13:13  & 138 & $-$48 & 132 & d\\
Triton & MMT & 2010 Jul 13   & 10:46  & 340 & $-$46 & 48 & c\\
\enddata
\tablenotetext{a}{Right-hand rule coordinates for Pluto}
\tablenotetext{b}{\cite{teg10}}
\tablenotetext{c}{This work}
\tablenotetext{d}{\cite{gru10}}
\end{deluxetable}

\clearpage

\begin{deluxetable}{ccccc}
\tablewidth{0pt}
\tablecaption{Positions of Pure CH$_4$ Bands and Shifts For CH$_4$ Diluted In N$_2$$^a$}
\tablehead{
\colhead{ Mode} & 
\colhead{$\lambda$} & 
\colhead{$\Delta$} &
\colhead{$\nu$} &
\colhead{$\Delta$} 
\\
\colhead{} &
\colhead{($\mu$m)} &
\colhead{($\mu$m)} &
\colhead{(cm$^{-1}$)} &
\colhead{(cm$^{-1}$)}}
\startdata
$\nu_1+2\nu_3+2\nu_4$  & 0.8897 & 0.0026 &11240.1 & 33.4 \\
$2\nu_3+2\nu_4$               & 1.1645 & 0.0032 & 8587.4 & 23.3 \\
$\nu_2+2\nu_3$                 & 1.3355 & 0.0034 & 7487.6 & 18.8 \\
$\nu_2+\nu_3+\nu_4$       & 1.7245 & 0.0048 & 5798.7 & 16.0 \\
$\nu_3+2\nu_4$                 & 1.7968 & 0.0047 & 5565.6 & 14.5 \\
$\nu_2+\nu_3$                   & 2.2081 & 0.0057 &4528.8 & 11.6 \\
\enddata
\tablenotetext{a}{From \cite{qui97}}
\end{deluxetable}

\clearpage

\begin{deluxetable}{ccccc}
\tablewidth{0pt}
\tablecaption{Methane Abundances For Pluto}
\tablehead{
\colhead{ Band} & 
\colhead{\%CH$_4$} &
\colhead{\%CH$_4$} &
\colhead{\%CH$_4$} &
\colhead{\%CH$_4$}
\\
\colhead{($\mu$m)} & 
\colhead{(10$^{\circ}$)} &
\colhead{(125$^{\circ}$)} &
\colhead{(194$^{\circ}$)} &
\colhead{(257$^{\circ}$)}}
\startdata
0.8897 & & & 7.9$^a$  &\\
1.1645 & 9.7 & 6.9 & & 8.8 \\
1.3355 & 9.2 & 7.2 & & 7.9 \\
1.7245 & 8.8 & 7.6 & & 8.1 \\
1.7968 & 8.5 & 6.9 & & 8.3 \\
2.2081 & 9.2 & 6.7 & & 8.1\\
avg & 9.1 & 7.1 & & 8.2 \\
std  & 0.5 & 0.4 & & 0.3 \\
\enddata
\tablenotetext{a}{Bok Observations, \cite{teg10}} 
\end{deluxetable}
\clearpage

\begin{deluxetable}{cccc}
\tablewidth{0pt}
\tablecaption{Methane Abundances For Triton}
\tablehead{
\colhead{ Band} & 
\colhead{\%CH$_4$} &
\colhead{\%CH$_4$} &
\colhead{\%CH$_4$}
\\
\colhead{($\mu$m)} & 
\colhead{(138$^{\circ}$)} &
\colhead{(314$^{\circ}$)} &
\colhead{(340$^{\circ}$)}}
\startdata
0.8897 &         & & 5.0$^a$ \\
1.1645 & 5.0  & 5.2& \\
1.3355 &  5.0 & 5.0  &\\
1.7245 &  5.0 & 5.5 &  \\
1.7968 & 5.2 & 6.0 & \\
2.2081 & 5.0 & 5.0  & \\
avg       & 5.0 & 5.3 & \\
std        & 0.1 & 0.4 & \\
\enddata
\tablenotetext{a}{MMT Observations, This Work} 
\end{deluxetable}
\clearpage

\begin{deluxetable}{ccrc}
\tablewidth{0pt}
\tablecaption{Methane Abundances For Eris}
\tablehead{
\colhead{} & 
\colhead{Band} &
\colhead{\%CH$_4$} &
\colhead{}
\\
\colhead{} & 
\colhead{($\mu$m)} &
\colhead{} &
\colhead{}}
\startdata
& 0.7291 & 8$^a$ & \\
& 0.8691 & 9$^a$ & \\
& 0.8897 & 10$^a$ & \\
& 1.7245 & 8$^b$ & \\
& 1.7968 & 14$^b$ & \\
& 2.2081 & 9$^b$ & \\
& avg  & 10 &  \\
& std     & 2 &  \\
\enddata
\tablenotetext{a}{Spectrum and analysis from \cite{teg10}} 
\tablenotetext{b}{Spectrum from \cite{dum07}; analysis from this work} 
\end{deluxetable}
\clearpage

\end{document}